\documentclass[reprint, aps,prl,twocolumn,showpacs,showkeys, 10pt, longbibliography, nolinenumbers, floatfix]{revtex4-2}

\usepackage{amsmath}
\usepackage{graphicx}
\usepackage{float}
\usepackage{dcolumn}
\usepackage{multirow}
\usepackage{natbib}
\usepackage{xcolor}
\usepackage[colorlinks = true,linkcolor = blue,urlcolor  = blue,citecolor = blue,anchorcolor = blue]{hyperref}
\usepackage{enumerate}

\begin{document}

\title{Constraints on the  isovector properties of finite nuclei from neutron stars observations   }

\author{M. Divaris$^1$}
\author{A. Kanakis-Pegios$^1$}
\author{Ch.C. Moustakidis$^1$}


\affiliation{$^1$Department of Theoretical Physics, Aristotle University of Thessaloniki, 54124 Thessaloniki, Greece  }

\begin{abstract}
The nuclear symmetry energy plays important role on the structure of finite nuclei as well as on the bulk properties of neutron stars.
However, its values at high densities are completely uncertain and the corresponding experimental data have a large error. One possibility to determine or at least estimate the values at high densities is with the help of neutron star observations. Recently, observations of gravitational waves from merging processes of binary neutron star systems provide useful information on both their radius and tidal deformability, quantities directly related to the symmetry energy. In this work, an attempt is made in this direction, namely to see how recent observations can help to constrain the structure of finite nuclei. In particular, in the present study we parameterize the equation of state which describes the asymmetric and symmetric nuclear mater with the help of the parameter $\eta=(K_0 L^2)^{1/3}$, where $K_0$ is the incompressibility and $L$ the slope parameter. The parameter $\eta$  is a regulator of the stiffness of the equation of state. We expect that the values of $\eta$ affect both the properties of finite nuclei as well as of the neutron star properties (where the role of the isovector  interaction plays important role). It is natural to expect that constraints, via the parameter $\eta$  on finite nuclei will imply constraints on the neutron star properties and vice versa. In view of the above statements we propose a simple but  self-consistent method  to examine simultaneously  the effects of the parameter $\eta$ on the properties of finite nuclei and neutron stars. We found constraints on the latter systems  via combination by the recent experiments (PREX-2) and observational data found by the detectors Ligo and Virgo.


\end{abstract}

\maketitle

\section{Introduction}

The Nuclear Symmetry Energy (NSE)  is one of the most fundamental quantities relevant to the study of both neutron-rich finite nuclei and neutron stars (for a comprehensive review see Refs.~\cite{Baldo-2016,Steiner-2005,Sammarruca-2013,Lattimer-2023,Lattimer-2014,BaoLI-2014a,Horowitz-2014,Bao-2021,Bao-2022}). This quantity is directly related to the isovector character of the nuclear forces and  exhibits a strong dependence on the baryonic density.
The uncertainty that exists in the knowledge of the symmetry energy, especially at low densities of nuclear matter, similar to those found in finite nuclei, can be partly addressed by terrestrial  experiments. However, its values at high densities are completely uncertain and the corresponding empirical data have a large error.

Both theoretical~\cite{Danielewicz-03,Danielewicz-09,Danielewicz-13,Lattimer-013,Piekarewicz-09,Moller-012,Ono-03,Brown-2000,
Brown-2001,Centelles-07,Centelles-09,Centelles-10,Warda-09,Vinas-03,
LWChen-05,Furnstahl-02,Kanzawa-09,Oyamatsu-07,Sammarruca-09,Agrawal-010,Agrawal-012,
LWChen-011,Mei-012,JLiu-013,Fattoyev-012,Fattoyev-013a,Zhang-013,Kortelainen-013,Singh-013,Paar-013,Mekjiian-85,
Bodmer-03,Denisov-02,Prassa-010,Wolter-09,Gaidarov-011,Gaidarov-012,
Reinhard-010,Sharma-09,Blocki-013,Erler-013,Moustakidis-07,
Moustakidis-012,Moustakidis-015,Paar-2014,Psonis-07,Fan-014,Papazoglou-2014,Inakura-013,Cozma-013,Fattoyev-013b,Mallik-013,Xu-013,Steiner-012,Souza-09,Drischler-014,Ravlic-2023,Lopes-2023,Patra-2023,Malik-2022,Cao-2022,Most-2021,Gaidarov-2021,Gil-2021,Hai-2020,Danchen-2020,Hai-2019,Raduta-2018,Maza-2018,Yong-2017,Mondal-2015,Antonov-2016,Alam-2016,Typel-2014,Deibel-2014,Lee-2014,Song-2015,Zhang-2014,Naz-2019,Kaur-2016,Goudarzi-2018,Gaidarov-2020,Tagami-2022,Sotani-2014,Sotani-2022,Guo-2023,Li-2019b,Nai-2019,Li-2021,Bertulani-2019,Dorso-2019,Pradhan-2023}
and experimental efforts~\cite{Tsang-09,Tsang-012,Klimkiewicz-07,Abrahamyan-012,Horowitz-012,
Tarbert-013,Trzcinska-01,Shetty-04,Shetty-07,Marini-013,Veselsky-013,Neil-2023} are focused on the study of a possible
correlation of the two main parameters concerning the NSE, that is the slope parameter   $L$ and its value at the saturation density of nuclear matter $J$, with various nuclear properties. These properties include mainly nuclear masses,  the neutron skin thickness, the nuclear dipole polarizability, the giant and pygmy dipole resonance  energies, flows in heavy-ion collisions and isobaric analog states (for a comprehensive analysis see Ref.~\cite{Lattimer-2023}). Moreover, there is  a variety of neutron star properties
that are sensitive to NSE   e.g. the radius and the maximum mass, the crust-core transition density and consequently  the  crust's  thickness, the thermal relaxation time, the various neutrino processes related with the cooling, and reaction rates involved in the astrophysical r-process~\cite{Lattimer-2023}. 

The vast majority of existing  equations of state   have been developed either to describe finite nuclei or to study the structure of neutron stars. There are rare cases that these equations have been used simultaneously to study these two systems in a self-consistent way. But when this is done then one can find a direct dependence of the microscopic properties related to the structure of finite nuclei with some of the macroscopic properties of neutron stars. And this is because the origin of their properties is common. Of course, it is also possible to use a variety of different approaches to nuclei and neutron stars (which will involve different methods and nuclear models) and then seek in a systematic study to correlate the properties of nuclei and neutron stars. In practice this is the method that has been used more widely in research (see the aforementioned references). 

The main motivation of the present study is to study in a self-consistent way, i.e. using the same nuclear model, both few of the  isovector  properties of finite nuclei and  of  neutron stars. The present work is based in our  previous study~\cite{Papazoglou-2014}, as far as the applied nuclear model for the description of finite nuclei is concerned.
However, we extend it in order to be suitable for the study of neutron stars.

In  particular, inspired by the previous study developed in Refs.~\cite{Sotani-2014,Sotani-2022},  we  parameterize the equation of state (EoS) which describes the asymmetric and symmetric nuclear matter with the help of the parameter $\eta=(K_0 L^2)^{1/3}$, where $K_0$ is the incompressibility and $L$ the slope parameter. The parameter $\eta$  is a regulator of the stiffness of the equation of state. 

The computational procedure we apply can be summarized as follows:
Firstly, 
based on the aforementioned parametrization, 
 we constructed a self-consistent
and easily applicable density functional method to study the
effects of the symmetry energy on the isovector structure
properties of medium and heavy neutron-rich nuclei. Secondly, we use these equations of state in order to study the structure and bulk properties of neutron stars. The advantage of this method is that one can study both finite nuclei and neutron stars with the same energy density functional. The key parameter $ \eta $ is essentially the {\it bridge} that connects the microscopic properties of nuclei to the macroscopic properties of NSs. Obviously possible experimental limitations on the properties of FN will be reflected in corresponding properties of NSs and vice versa.

In this paper we  use the data from the PREX-2 experiment concerning  the neutron skin thickness of $^{208}$Pb~\cite{Adhikari-2021,Reed-2021}. In this experiment the thickness values are quite large compared to other corresponding experiments and this places strong constraints on the slope of the nuclear symmetry energy  demanding a stiff equation of state at least for densities close to the saturation density. Moreover, we use observational
constraints from the GW170817 event~\cite{Abbott-2019-X}  concerning the tidal deformability of neutron stars.
These observations lead in general to a softer equation of state. Therefore, these conflicting results, can lead to strong constraints on the symmetry energy and hence the nuclear matter equation of state. In the present study we will show how the above constraints can improve our knowledge both in the structure of finite nuclei and neutron stars. 

The paper is organized as follows: in Section 2, we present the theoretical model for the study both of finite nuclei and neutron stars. Section 3, is dedicated to the presentation of the results and to relevant discussion. Finally, in Section 4, we finalize our investigation with the concluding remarks.

\section{The theoretical nuclear model}
The key quantity in our calculations is  the energy per particle of asymmetric nuclear matter, where in a  good approximation, at  least for densities close to the saturation density, is given by the expression~\cite{Lattimer-2023,Lattimer-2014, Piekarewicz-09}
\begin{equation}
 E(n,\alpha)=E_0+\frac{K_0}{18 n_0^2}  \left(n-n_0\right)^2+ S(n)\alpha^2
 \label{En-per-bar}
\end{equation}
where $\alpha=(n_n-n_p)/n$ is the asymmetry parameter, with $n_n$ and $n_p$ the number densities of neutrons and protons respectively and   $n_0$ is the  saturation density. Moreover $E_0=E(n_0,0)$ is the energy per particle at  $n_0$,  $K_0$ is the  incompressibility and $S(n)$ is the symmetry energy.  In particular, the nuclear symmetry energy $S(n)$ can be developed in a series around the saturation density
\begin{equation}
S(n)=J+\frac{L}{3n_0} (n-n_0)+\frac{K_{\rm sym}}{18n_0^2} (n-n_0)^2+\cdots
 \label{Sym-1}   
\end{equation}
where  $J=S(n_0)$.  The slope parameter $L$ is related to  the first derivative  and  $K_{\rm sym}$ to the second derivative of the NSE according to the definitions 
\begin{equation}
 L= 3n_0\left (\frac{dE_{\rm sym}(n)}{dn}  \right)_{n=n_0}
 \label{L-1}
\end{equation}
\begin{equation}
 K_{\rm sym}= 9n_0^2\left (\frac{dE_{\rm sym}^2(n)}{d^2n}  \right)_{n=n_0}
 \label{K0-1}
\end{equation}
In the present work, we will omit the third term in the expansion~(\ref{Sym-1}) which has a small contribution compared to the others. Now the  
 corresponding energy density ${\cal E}=nE$, which is the key quantity in the present study and essentially serves to {\it bridge} the properties of finite nuclei and nuclear matter (neutron star matter) reads 
\begin{eqnarray}
 {\cal E}_b(n,\alpha) &=&E_0 n+\frac{K_0}{18 n_0^2}n  \left(n-n_0\right)^2 \nonumber \\
 &+&\left(J+\frac{L}{3n_0}
 (n-n_0)\right)n\alpha^2 
 \label{EnDen-per-bar}
\end{eqnarray}
Expression (\ref{EnDen-per-bar}) is a very good approximation for densities close to the saturation density but can be extended also to higher values of densities. This expression  will be used  to infer the properties of finite nuclei, mainly focus on those related with the isovector character of the nuclear forces, as well as the bulk neutron star properties including mainly the mass, radius and tidal deformability.

\subsection{Finite Nuclei}
According to the empirical Bethe-Weizsacker formula the
binding energy of a finite nucleus with $A$ nucleons and atomic
number $Z$ is given by 
\begin{eqnarray}
 BE(A,Z)&=&-a_V A+a_SA^{2/3}+a_C\frac{Z(Z-1)}{A^{1/3}}   \nonumber \\
 &+&a_A\frac{(N-Z)^2}{A}+E_{\rm add}
 \label{BW-1}
\end{eqnarray}
The first term corresponds to the volume effect, the second is
the surface term, the third one takes into account the Coulomb
repulsion of the protons, while the fourth is the symmetry
energy term. Finally, the last term $E_{\rm add}$ corresponds to other additional factors including the pairing interaction. Using fits of known masses to
this equation one can determine the corresponding coefficients
$a_V$, $a_S$, $a_C$, and $a_A$.

Each term of the  Bethe-Weizsacker formula can be derived by employing the density functional theory. In this case, the total energy is a functional
of the proton and neutron densities and consists of terms
corresponding with those appearing in formula~(\ref{BW-1}). The
minimization of the total energy functional  defines the related densities
and consequently the contribution of each term separately.

For finite nuclei, we consider the total energy of the nucleus in terms of an energy density functional of the proton $ \rho_{p} (r) $ and neutron $ \rho_{n} (r) $ number densities
\begin{equation}
E =  \int_{\cal{V}}  {\cal E} \left( \rho ( r ), \alpha ( r ) \right)  d^3 r
\label{funct-1}
\end{equation}
where $ {\cal E}({\rho} (r),{\alpha} (r)) $ is the local energy density, $ \rho = \rho_{n} + \rho_{p} $ is the total number density and $ \alpha = (\rho_n - \rho_p)/(\rho_n + \rho_p) $ is the asymmetry function. The integration is performed over the total volume $ \cal V $ of the nucleus. In the present work we consider the functional
\begin{equation}
E =  \int_{\cal{V}} \left( {\cal E}_b (\rho, \alpha) 
+ F_o |\nabla \rho(r)|^2 + \frac{1}{4}  \rho (1-a) V_C (r) \right) d^3 r
\label{funct-2}
\end{equation}
where $ {\cal E}_b $ is the energy density of asymmetric nuclear matter, the second term is the gradient term originating from the finite-size character of the density distribution and the third term is the Coulomb energy density. The Coulomb potential is given by 
\begin{equation}
V_C (r) = \int_{\cal V} \frac{e^2 \rho_{p} (r')}{ |\textbf{r}-\textbf{r}'| } d^3 r' = \\
\frac {e^2}{2} \int_{\cal V} \frac{\rho (1- \alpha (r'))}{|\textbf{r}-\textbf{r}'|} d^3 r'
\label{Vc}
\end{equation}
The Poisson equation for the Coulomb potential $ {\nabla ^2} V_C (r)= - 4 \pi e^2 {\rho_p} (r) $ can be used to check the convergence of the iteration process involved in the calculations.
The total number density $ \rho (r) $ and the asymmetry function $ \alpha (r) $ obey respectively the following constraints:
\begin{equation}
\int_{\cal V} \rho (r) d^3 r = A
\label{norm-rho}
\end{equation}
\begin{equation}
\int_{\cal V} \alpha (r) \rho (r) d^3 r = N-Z
\label{norm-alpha}
\end{equation}
The minimization of the total energy, given by the functional~(\ref{funct-2}), with the above constraints constitutes a variational problem. Defining 
\begin{equation}
h= 4 \pi r^2 ( {\cal E} + {\lambda_1} \rho (r) + {\lambda_2} \alpha (r) \rho (r) )
\end{equation}
the total energy E obtains a minimum for the solutions of the following differential equations
\begin{equation}
\frac{\partial h}{\partial \rho} - \frac {d}{dr} \frac{\partial h}{\partial \rho '}=0
\end{equation}
\begin{equation}
\frac{\partial h}{\partial \alpha} - \frac {d}{dr} \frac{\partial h}{\partial \alpha '}=0
\end{equation}
From the first differential equation, a second order differential equation for $ \rho (r) $ is extracted
\begin{equation}
2 F_o \frac{d^2 \rho}{dr^2} + \frac {4 F_o}{r} \frac{d \rho}{dr} - \\
\frac{\partial {\cal E}_b }{\partial \rho} - \\
\frac{1}{4} (1- \alpha)V_C - {\lambda_1} - \alpha {\lambda_2}=0
\label{diff-eq-rho}
\end{equation}
while from the second, we obtain an equation that provides us with the Lagrange multiplier $ {\lambda_2} $
\begin{equation}
\frac{\partial {\cal E}_b }{\partial \alpha} - \frac{1}{4}  \rho V_C + \\
{\lambda_2} \rho (r) =0
\label{diff-eq-alpha}
\end{equation}
which gives
\begin{equation}
\alpha (r) =\frac{V_C}{8 S(\rho)}- \frac{\lambda_2}{2 S(\rho)}
\label{alpha-2}
\end{equation}
One has to solve self-consistently the above system of differential equations in order to extract the total number density $ \rho (r) $.  In the present work, in order to avoid the complication due to the differential
equation (\ref{diff-eq-rho}) we employ a variational method where use is
made of an appropriate trial function for $\rho(r)$ (see Refs.~\cite{Brueckner-68,Brueckner-69,Brueckner-71,Buchler-71,Lombard-73}). This method,
provides a convenient tool in seeking an approximate solution
for heavy nuclei. 
In the present study we use a Fermi-type trial density function of the form
\begin{equation}
\rho (r)= \frac{\rho_0 }{1+{\rm exp}[(r-d)/w] }
\end{equation}
The asymmetry function $\alpha(r)$ obeys the constraints $0\leq
\alpha(r) \leq 1$. However, expression (\ref{alpha-2}) does
not ensure the above constraints, since for high values of $r$
(low values of $\rho(r)$ and consequently $S(\rho)$) $\alpha(r)$
increases very fast and there is a cut-off radius, $r_c$ where
$\alpha(r_c)=1$ and also $\alpha(r \geq r_c)\geq 1$.
In order to overcome this unphysical behavior of $\alpha(r)$ we
use the assumption
\begin{eqnarray}
\alpha(r)&=&\left\{
\begin{array}{ll}
\frac{1}{8S(\rho)}\left(\frac{}{} V_c(r)-4\lambda_2
\right), \qquad r \leq r_c       &          \\
\\
1, \qquad r\geq r_c.  &  \
                              \end{array}
                       \right.
\label{resip-2}
\end{eqnarray}
%
%
%
%
Now, one  possibility is  to calculate the symmetry energy coefficient
$a_{A}$, defined in the Bethe-Weizsacker formula,  via the local
density approximation. In this approach $a_A$ is defined by the
integral
\begin{equation}
a_A=\frac{A}{(N-Z)^2}\int_{\cal V}\rho(r) S(\rho)\alpha^2(r) d^3r.
\label{sym-coef-1}
\end{equation}
Definition (\ref{sym-coef-1}) shows explicitly the direct strong
dependence of $a_A$ on  the symmetry energy $S(\rho)$ and the
asymmetry function $\alpha(r)$. Actually, according to the present
study,  the total integral is split in two parts as follows
\begin{eqnarray}
a_A&=&\frac{A}{(N-Z)^2}\left(\int_{{\cal V}_c}\rho(r)
S(\rho)\alpha^2(r) d^3r \right.\nonumber \\
&+&\left.\int_{{\cal V}_{\rm shell}}\rho(r) S(\rho)
d^3r\right). \label{resip-24}
\end{eqnarray}
where ${\cal V}_c$ is the spherical volume corresponding  to the radius $r_c$ and ${\cal V}_{\rm shell}$  is the spherical shell bounded internally by the radius  $r_c$ and externally by the radius of the nucleus. 

It was suggested that the symmetry energy coefficient $a_A$ can be expanded as 
determined by the formula \cite{Danielewicz-13}
\begin{equation}
a_A^{-1}=(a_A^V)^{-1}+(a_A^S)^{-1}A^{-1/3}.
\label{fit-Daniel}
\end{equation}
In the present work, we make use of a new expression for the asymmetry coefficient $ a_A $ given by
\begin{equation}
a_A =\frac{A}{{\cal I}_1}\left(1+ \Delta_C \right)
\label{aA-new}
\end{equation}
where the quantity $\Delta_C$ reads
\begin{eqnarray}
\Delta_C &=& \frac{1}{64(N-Z)^2}\biggl({\cal I}_1\left({\cal I}_3+64{\cal I}_5\right) \nonumber \\
&-& {\cal I}_2^2 + 64{\cal I}_4^2 -128{\cal I}_4 (N-Z) \biggr)
\label{DeltaC}
\end{eqnarray}
with
\begin{eqnarray}
{\cal I}_1&=&\frac{A}{J}+\frac{1}{J}\int_{{\cal V}} \rho(r)\left(\frac{J}{S(\rho)}-1 \right) d^3r \nonumber \\
&-& \int_{{\cal V}_{\rm shell}} \frac{\rho(r)}{S(\rho)} d^3 r =\frac{A}{J} + \frac{1}{J} {\cal I}_6 - {\cal I}_7
\label{A-f-3}
\end{eqnarray}
\begin{equation}
{\cal I}_2=\int_{{\cal V}_c} \frac{V_c(r)\rho (r)} {S(\rho)} d^3 r 
\label{I-2}
\end{equation}
\begin{equation}
{\cal I}_3=\int_{{\cal V}_c} \frac{V^2_c(r)\rho (r)} {S(\rho)} d^3 r 
\label{I-3}
\end{equation}
The integrals $ {\cal I}_4 $ and $ {\cal I}_5 $ are given by
\begin{equation}
{\cal I}_4=\int_{{\cal V}_{\rm shell}} \rho (r) d^3 r ,\quad {\cal I}_5=\int_{{\cal V}_{\rm shell}} \rho (r) S(\rho) d^3 r
\label{integrals-4-5}
\end{equation}
and also 
\begin{eqnarray}
{\cal I}_6 &=& \int_{{\cal V}} \rho (r)\left( \frac{J}{S(\rho)} -1 \right) d^3 r ,\nonumber \\ 
{\cal I}_7 &=& \int_{{\cal V}_{\rm shell}} \frac{\rho (r)} {S(\rho)} d^3 r
\label{integrals-6-7}
\end{eqnarray}
Finally, we obtain for the volume $ a_A^V $ and surface $a_A^S $ contributions to the asymmetry coefficient $a_A$, which appear in (\ref{fit-Daniel}), the expressions
\begin{equation}
a_A^V = J(1+  \Delta_C), \quad a_A^S = Q_S (1+ \Delta_C)
\label{aAV-aAS}
\end{equation}
where
\begin{equation}
Q_S=\frac{J A^{2/3}}{{\cal I}_6 - J{\cal I}_7}
\label{QS}
\end{equation}
One of the most important quantities concerning the isovector
character of the nuclear forces is the neutron skin  thickness
defined as
\begin{equation}
\Delta R_{\rm skin} = R_n - R_p
\end{equation}
with
\begin{equation}
R_n = \left( \frac{1}{N} \int_{\cal V} r^2 \rho_n d^3 r \right)^{1/2}=\left( \frac{1}{N} \int_{\cal V} r^2 \frac{\rho (1+\alpha)}{2} d^3 r \right)^{1/2}
\label{Rn}
\end{equation}
and
\begin{equation}
R_p = \left( \frac{1}{Z} \int_{\cal V} r^2 \rho_p d^3 r \right)^{1/2}=\left( \frac{1}{Z} \int_{\cal V} r^2 \frac{\rho (1-\alpha)}{2} d^3 r \right)^{1/2}
\label{Rp}
\end{equation}
It is worth mentioning that $\Delta R_{\rm skin}$ is not directly dependent on $S(\rho)$,
compared to the case of $a_A$. However, it is dependent  indirectly via the asymmetry function  $\alpha(r)$. Thus it is reasonable to expect  $\Delta R_{\rm skin}$, as well as the coefficients $a_A$, $a_A^S$, and $a_A^V$ to be  strong indicators of the isospin character of the nuclear interaction.


\subsection{Neutron Stars}
The equation of state of neutron star matter is the key quantity to study the structure and the properties of neutron stars~\cite{Shapiro-1983,Haensel-2007,Bielich-2020}. It consists mainly by two parts. The first one is the contribution of the baryons (neutrons and protons mainly) and the second is the contribution by leptons (mainly electrons and muons). In the present work the contribution on the energy density of neutron star matter is given by the expression (\ref{EnDen-per-bar}). The pressure, due to the baryons which is  defined as 
\begin{equation}
P_b=n^2\frac{d({\cal E}/n)}{dn}
 \label{Pres-def}   
\end{equation}
reads now 
\begin{equation}
P_b=\frac{K_0}{9 n_0^2}n^2  \left(n-n_0\right)+\alpha^2 \frac{L}{3n_0}n^2
\label{Pres-1}    
\end{equation}
The contribution to the total energy density and pressure by electrons is given by the well known formula of the relativistic  Fermi gas, that is 
\begin{equation}
{\cal E}_e=\frac{(m_ec^2)^4}{8\pi^2 (\hbar c)^3} \left[z(2z^2+1)\sqrt{1+z^2}-\ln\left( z+\sqrt{z^2+1}\right)  \right]   
\label{Ed-ele}
\end{equation}
and 
\begin{equation}
P_e=\frac{(m_ec^2)^4}{24\pi^2 (\hbar c)^3} \left[z(2z^2-3)\sqrt{1+z^2}+3\ln\left( z+\sqrt{z^2+1}\right)   \right]   
\label{Pre-ele}
\end{equation}
\begin{equation}
z=\frac{(\hbar c) (3\pi^2 n_e)^{1/3}}{m_e c^2}, \quad n_e=x_p n  
\label{xelectron}
\end{equation}
where $x_p$ is the proton fraction. 
The total energy density and pressure of charge neutral and chemical equilibrium matter is 
\begin{equation}
 {\cal E}_{\rm tot}={\cal E}_b+{\cal E}_e
 \label{en-total}
\end{equation}
\begin{equation}
 P_{\rm tot}=P_b+P_e
 \label{Pres-total}
\end{equation}
From Eqs.~(\ref{en-total}) and (\ref{Pres-total}) we construct the equation of state of neutron star matter. 

The proton fraction $x_p=n_p/n$ which plays a crucial role on neutron star properties is a quantity very sensitive  on the NSE. In particular, the condition of beta equilibrium in the interior of neutron stars
\begin{equation}
\mu_n=\mu_p+\mu_e    
\end{equation}
where $\mu_i$ ($ i=n,p,e$) are the chemical potentials of protons, neutrons and electrons, leads to the following equation
\begin{equation}
 4(1-2x_p)S(n)=\hbar c (3\pi^2 n_e)^{1/3}=\hbar c (3\pi^2 n x_p)^{1/3}
 \label{cond-1}
\end{equation}
Solving Eq.~(\ref{cond-1}) we found the density dependence of the proton fraction
\begin{equation}
x_p(n)=\frac{1}{2}-\frac{1}{4}\left( [2\beta(\gamma+1)]^{1/3}- [2\beta(\gamma-1)]^{1/3}\right)
\label{sol-1}
\end{equation}
where
\[\beta=3\pi^2n\left(\frac{\hbar c}{4 S(n)}  \right)^3, \qquad \gamma=\left(1+\frac{2\beta}{27}  \right)^{1/2}  \]
Having now constructed the equation of state of neutron star matter we can calculate their basic properties by solving the Tolman-Oppenheimer-Volkoff equations that express the hydrostatic equilibrium.

\subsection{TOV equations and tidal deformability}
The mechanical equilibrium of the star matter is determined by the system of two differential equations, the
well known Tolman–Oppenheimer–Volkoff (TOV) equations and  the equation of state ${\cal E}={\cal E}(P)$ of the fluid.
This system reads
\begin{eqnarray}
\frac{dP(r)}{dr}&=&-\frac{G{\cal E}(r) M(r)}{c^2r^2}\left(1+\frac{P(r)}{{\cal E}(r)}\right) \nonumber \\
&\times&
 \left(1+\frac{4\pi P(r) r^3}{M(r)c^2}\right) \left(1-\frac{2GM(r)}{c^2r}\right)^{-1},
\label{TOV-1}
\end{eqnarray}
\begin{equation}
\frac{dM(r)}{dr}=\frac{4\pi r^2}{c^2}{\cal E}(r).
\label{TOV-2}
\end{equation}
The solving of the coupled differential equations (\ref{TOV-1}) and (\ref{TOV-2}) for $P(r)$ and $M(r)$ requires their numerical integration from the origin ($r = 0$) to the point $r=R$ where the pressure becomes practically zero. At this point the radius and the mass of the neutron star are computed. We notice that each EoS leads to an infinite number of configurations to each of which corresponds a pair of mass and radius. What is mainly of interest, in any case, is the predicted maximum mass and radius corresponding to a mass equal to $1.4$ solar masses. In particular the maximum mass is related with the stiffness of the EoS and in any case must be larger than the already observed masses, so as to ensure the plausibility of the corresponding EoS.  Furthermore, the radius corresponding to 1.4 solar masses is already subject to many constraints derived from both observations and robust theoretical predictions and gives useful information for the EoS at low densities, close to the saturation density.

The last years very useful information has been obtained from observations of gravitational waves resulting from the merger of black hole–neutron star and neutron star–neutron star binary systems. We notice that this kind of source leads to the measurement of various properties of neutron stars. During the inspiral phase of the binary neutron star systems, the tidal effects can be detected. To be more specific, the tidal Love number $k_2$ describes the response of the neutron star to the tidal field and depends both on the neutron star mass and the applied EoS. The exact relation which describes the tidal effects is given below~\cite{Flanagan-08,Hinderer-08}
\begin{equation}
Q_{ij}=-\frac{2}{3}k_2\frac{R^5}{G}E_{ij}\equiv- \lambda E_{ij},
\label{Love-1}
\end{equation}
where $\lambda$ is the tidal deformability. The tidal Love number $k_2$ is given by \cite{Flanagan-08,Hinderer-08}
\begin{eqnarray}
k_2&=&\frac{8\beta^5}{5}\left(1-2\beta\right)^2\left[2-y_R+(y_R-1)2\beta \right]\nonumber\\
& \times&
\left[\frac{}{} 2\beta \left(6  -3y_R+3\beta (5y_R-8)\right) \right. \nonumber \\
&+& 4\beta^3 \left.  \left(13-11y_R+\beta(3y_R-2)+2\beta^2(1+y_R)\right)\frac{}{} \right.\nonumber \\
&+& \left. 3\left(1-2\beta \right)^2\left[2-y_R+2\beta(y_R-1)\right] {\rm ln}\left(1-2\beta\right)\right]^{-1}_,
\label{k2-def}
\end{eqnarray}
where $\beta=GM/Rc^2$ is the compactness of a neutron star. The parameter $y_R$ is determined by the following differential equation~\cite{Flanagan-08,Hinderer-08}
\begin{equation}
r\frac{dy(r)}{dr}+y^2(r)+y(r)F(r)+r^2Q(r)=0 
\label{D-y-1}
\end{equation}
$F(r)$ and $Q(r)$ are functions of the energy density ${\cal E}(r)$, pressure $P(r)$, and mass $M(r)$ defined as
\begin{equation}
F(r)=\left[ 1- \frac{4\pi r^2 G}{c^4}\left({\cal E} (r)-P(r) \right)\right]\left(1-\frac{2M(r)G}{rc^2}  \right)^{-1},
\label{Fr-1}
\end{equation}
and
\begin{eqnarray}
r^2Q(r)&=&\frac{4\pi r^2 G}{c^4} \left[5{\cal E} (r)+9P(r)+\frac{{\cal E} (r)+P(r)}{\partial P(r)/\partial{\cal E} (r)}\right]
\nonumber\\
&\times&
\left(1-\frac{2M(r)G}{rc^2}  \right)^{-1}- 6\left(1-\frac{2M(r)G}{rc^2}  \right)^{-1} \nonumber \\
&-&\frac{4M^2(r)G^2}{r^2c^4}\left(1+\frac{4\pi r^3 P(r)}{M(r)c^2}   \right)^2\left(1-\frac{2M(r)G}{rc^2}  \right)^{-2}.
\label{Qr-1}
\end{eqnarray}
Eq.\textcolor{blue}{(~\ref{D-y-1})} must be solved numerically and self consistently with the TOV equations under the following boundary conditions: $y(0)=2$, $P(0)=P_c$ ($P_{c}$ denotes the central pressure), and $M(0)=0$. The numerical integration provides the value of $y_R=y(R)$, which is a basic ingredient for $k_2$.

In addition, an important and well measured quantity by the gravitational wave detectors, which can be treated as a tool to impose constraints on the EoS, is the dimensionless tidal deformability $\Lambda$, defined as 
\begin{equation}
    \Lambda=\frac{2}{3}k_2 \left(\frac{c^2R}{GM}\right)^5=\frac{2}{3}k_2 (1.473)^{-5}\left( \frac{R}{{\rm Km}} \right)^5\left(\frac{M_{\odot}}{M}  \right)^5
\end{equation}
We notice that $\Lambda$ is sensitive to the neutron star radius, hence can provide information for the low density part of the EoS, which is related also to the structore and properties of finite nuclei.


\section{Results and Discussion}
Firstly, we calculate the properties of the nucleus $^{208}$Pb using the functional given by Eq.~(\ref{funct-2}). In particular for various values of the parameter $\eta$ we calculate the neutron skin  $\Delta R_{\rm skin}$ and the coefficients $a_A$, $a_A^S$ and  $a_A^V$. The results are presented in  Table~(\ref{Table-1}). Obviously the effects of the stiffness of the EoS are more pronounced in the case of the skin and the coefficients $a_A$ and  $a_A^S$ and moderately for $a_A^V$. In particular, the comparison of the skin with the experimental  data from PREX-2 can lead  to some constraints on the parameter $\eta$. Considering that the values of the skin of $^{208}$Pb reported by PREX-2 are~\cite{Adhikari-2021,Reed-2021}
\begin{equation}
 \Delta R_{\rm skin}=(0.283\pm 0.071) \ {\rm fm}  
\end{equation}
where the quoted uncertainty represents a 1$\sigma$ error, we conclude that the values of $\eta$  are roughly limited in the interval $\eta\sim [110-120] \ {\rm MeV}$. Corresponding restrictions apply to the coefficients $a_A$, $a_A^S$ and  $a_A^V$. It is useful to recall here there are some empirical relationships for the values of the above coefficients, as reported in Ref.~\cite{Danielewicz-13}. More precisely, in this paper, the authors 
 using excitation energies to isobaric analog states (IAS) and charge invariance found that the dependence of the mass coefficient  $a_A$ (see Eq.(\ref{fit-Daniel}))  can be well described in terms of a macroscopic volume–surface competition formula
with $a_A^S\simeq 10.7$ MeV and  $a_A^V\simeq 33.2 $ MeV. These  values suggest that the appropriate interval for $\eta$ is the one mentioned above.

\begin{table}
\caption{The  incompressibility $K_0$ (in MeV), the slope parameter $L$ (in MeV), the parameter $\eta$ (in MeV),  the $\Delta R_{\rm skin}$ (in fm), $a_A$ (in MeV), $a_A^S$ (in MeV), $a_A^V$ (in MeV) correspond to the various equations of state.}
\begin{tabular}{ c c c c c c c }
\hline
$ K_0 $ & $ L $ & $ \eta $ &  $ \Delta R _{skin} $ & $ a_A $ & $ a_A^S $ & $ a_A^V $ \\	
\hline
220	& 40 & 70.61 & 0.0462 & 27.870 & 35.591 & 32.114 \\
\hline
224	& 48 & 80.21 & 0.0693 & 26.846 & 27.565 & 32.127 \\
\hline
228	& 56 & 89.42 & 0.0971 & 25.718 & 21.713 & 32.144 \\
\hline
232 & 64 & 98.31 & 0.1316 & 24.445 & 17.185 & 32.167 \\
\hline
236 & 72 & 106.95 & 0.1768 & 22.953 & 13.489 & 32.201  \\
\hline
240 & 80 & 115.38 & 0.2420 & 21.090 & 10.283 & 32.257  \\
\hline
244	& 88 & 123.63 & 0.3504 & 18.442 & 7.836 & 30.594 \\
\hline
248 & 96 & 131.72 & 0.4376 & 14.839 & 6.4756 & 24.198 \\
\hline
252 & 104 & 139.69 & 0.5054 & 10.477 & 4.576 & 17.075 \\
\hline
256	& 112 & 147.53 & 0.5624 & 5.4637 & 2.366 & 8.953 \\
\hline
\end{tabular}
\label{Table-1}
\end{table}

\begin{table}
\caption{The incompressibility $K_0$ (in MeV), the slope parameter $L$ (in MeV), the parameter $\eta$ (in MeV),  the $R_{\rm max}$ (in Km), $M_{\rm max}$ (in $M_{\odot}$), $R_{1.4}$ (in Km), $\Lambda_{\rm max}$ and $\Lambda_{1.4}$ correspond to the various equations of state.}
\begin{tabular}{ c c c c c c c c }
\hline
$ K_0 $ & $ L $ & $ \eta $ &  $ R_{max} $ & $ M_{max} $ & $ R_{1.4} $ & $\Lambda_{max} $ & $\Lambda_{1.4} $\\	
\hline
220	& 40 & 70.61 & 10.776	& 2.342 & 12.151 & 2.624 & 333.362 \\
\hline
224	& 48 & 80.21 & 10.898 & 2.356 & 12.386 & 2.752 & 384.534 \\
\hline
228	& 56 & 89.42 & 11.013 & 2.369 & 12.612 & 2.881 & 441.022 \\
\hline
232 & 64 & 98.31 & 11.124 & 2.381 & 12.848 & 3.008 & 505.311 \\
\hline
236 & 72 & 106.95 & 11.230 & 2.392 & 13.083 & 3.159 & 579.319 \\
\hline
240 & 80 & 115.38 & 11.336 & 2.403 & 13.331 & 3.353 & 664.153 \\
\hline
244	& 88 & 123.63 & 11.441 & 2.413 & 13.590 & 3.495 & 767.730 \\
\hline
248 & 96 & 131.72 & 11.546 & 2.423 & 13.874 & 3.643 & 895.909 \\
\hline
252 & 104 & 139.69 & 11.654 & 2.433 & 14.185 & 3.829 & 1048.289 \\
\hline
256	& 112 & 147.53 & 11.767 & 2.442 & 14.535 & 4.004 & 1252.559 \\
\hline
\end{tabular}
\label{Table-2}
\end{table}



%
\begin{figure}[ht]
\includegraphics[width=228pt,height=18pc]{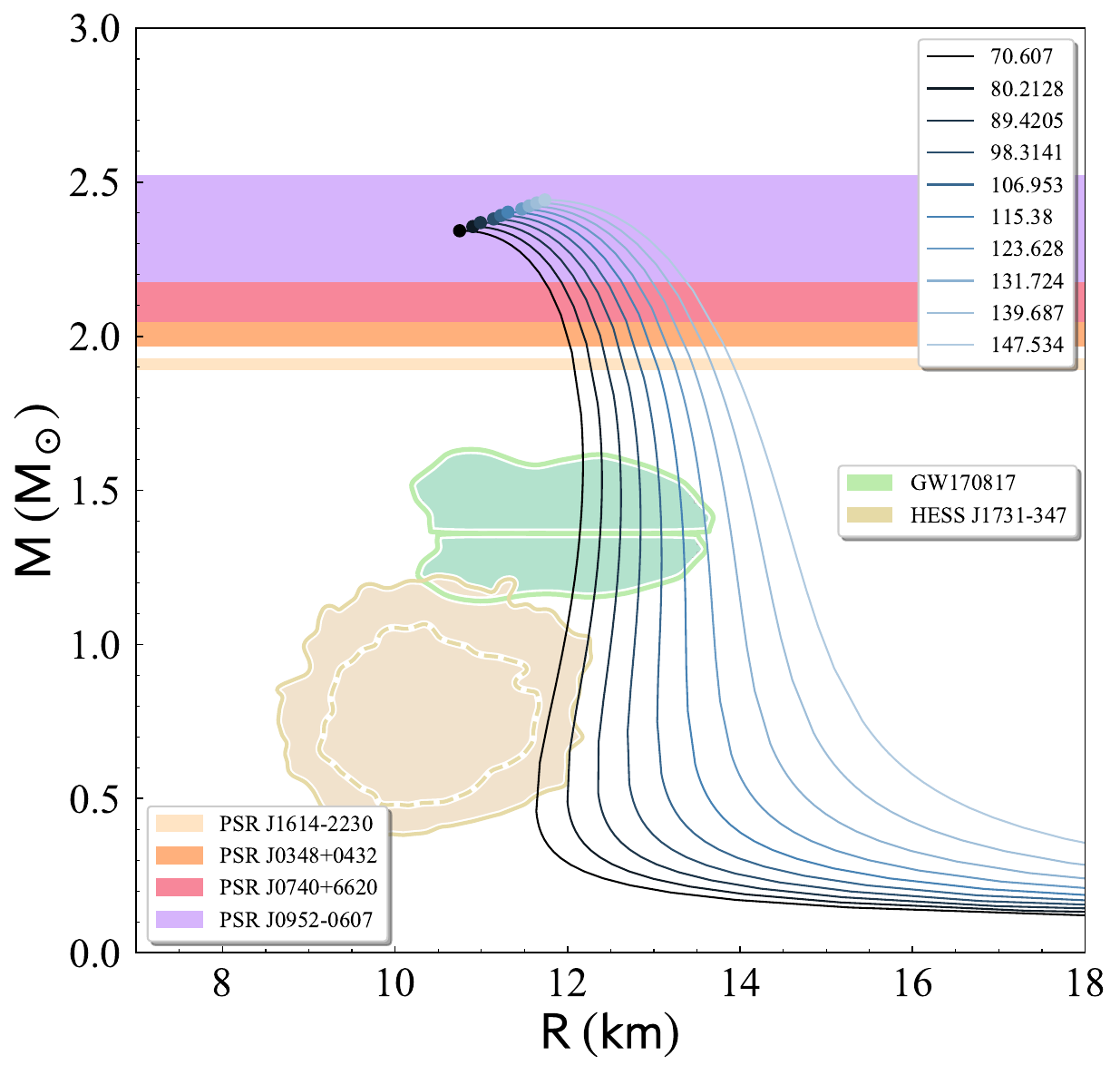}
\caption{The mass-radius (M-R) dependence for various EoSs depending on the parameter $\eta$. Various astrophysical constraints have been included for comparison (shaded regions). The shaded regions from bottom to top represent the HESS J1731-347 remnant~\cite{Doroshenko-2022}, the GW170817 event~\cite{Abbott-2019-X}, PSR J1614-2230~\cite{Arzoumanian-2018}, PSR J0348+0432~\cite{Antoniadis-2013}, PSR J0740+6620~\cite{Cromartie-2020}, 
and PSR J0952-0607~\cite{Romani-2022} pulsar observations for the possible maximum mass. }
\label{fig:M-R}
\end{figure}
\begin{figure}[ht]
\includegraphics[width=228pt,height=18pc]{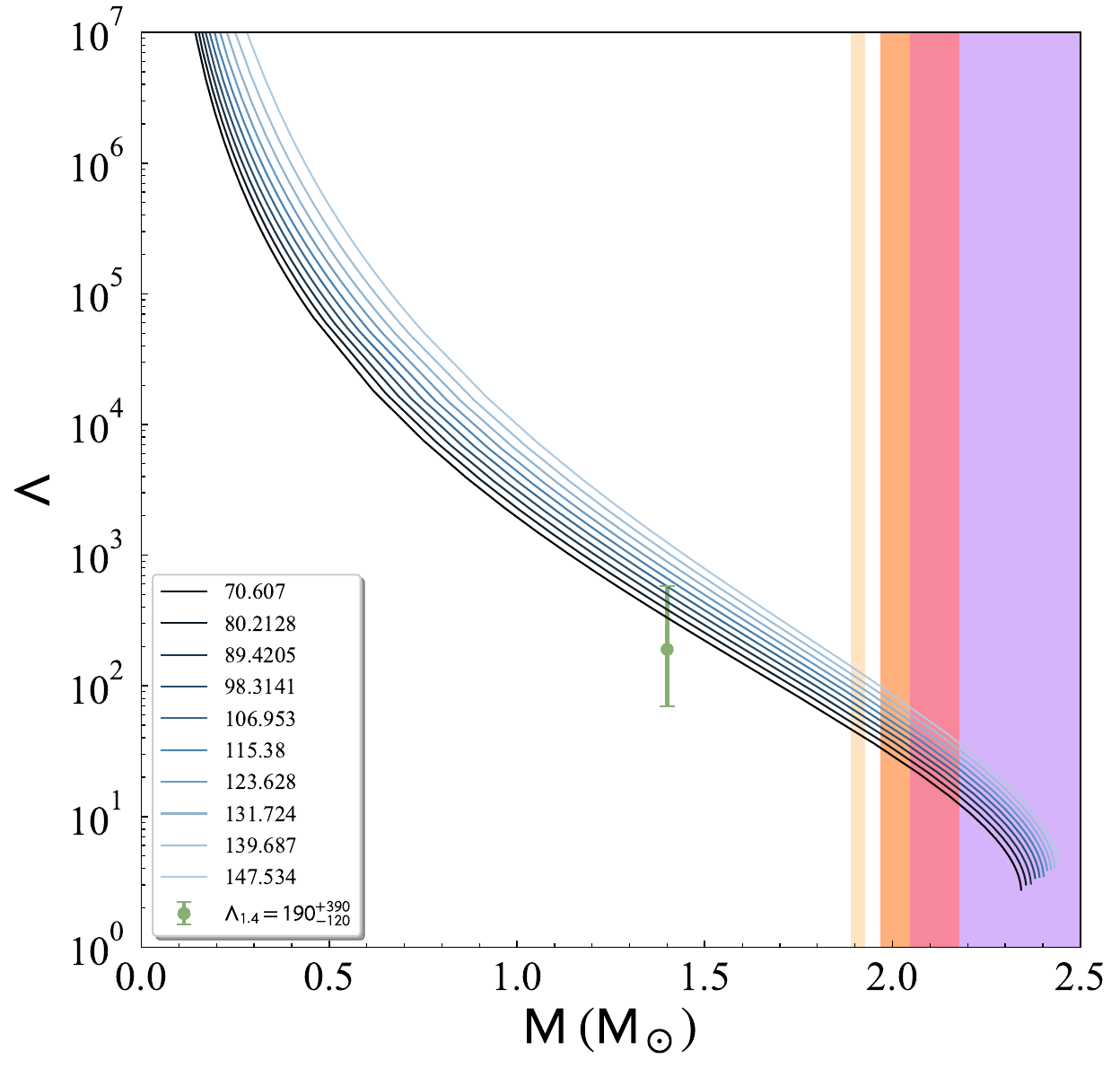}
\caption{The dimensional tidal deformability $\Lambda$ as a function of the mass for various EoSs corresponding to the  selected values of the parameter $\eta$. The shaded vertical regions indicate observational estimations~\cite{Abbott-2019-X}.}
\label{fig:Lambda-M}
\end{figure}

We also extend the study to include the basic properties of neutron stars. We focus on the same cases as for finite nuclei where we use the parameter $\eta$ as the key parameter. In particular, in Table (\ref{Table-2}) we provide the values of the maximum mass  $M_{\rm max}$ (in $M_{\odot}$), the corresponding radius 
$R_{\rm max}$ (in Km), the values of the radius $R_{1.4}$ (in Km) which correspond to mass $1.4 \ M_{\odot}$ the tidal deformability  corresponding to the maximum mass  $\Lambda_{\rm max}$ and finally the tidal deformability $\Lambda_{1.4}$ corresponding to mass $1.4 \ M_{\odot}$,  for each specific case.

In Fig.~\ref{fig:M-R} we show the relation between the mass and the radius of a single neutron star. The EoSs characterized by the parameter $\eta$ are shown with solid curves, with the lighter colors corresponding to higher values of $\eta$. The shaded regions indicate the observational data of different origin. In general, all the EoSs predict a high value for the maximum mass $\mathrm{M_{max}}$, with $\mathrm{M_{max}}$  increasing as the value of $\eta$ grows. In addition, the radius increases accordingly to $\eta$. In general, the increasing of $\eta$ affects more the radius compared to the $\mathrm{M_{max}}$. The EoSs with the highest $\eta$ lie outside of the GW170817 observation~\cite{Abbott-2019-X} (green shaded contours), while only the EoS with the lowest value of $\eta$ can predict the HESS observation~\cite{Doroshenko-2022}.

The dimensionless tidal deformability related to the mass of a single neutron star for all EoSs that we used is shown in Fig.~\ref{fig:Lambda-M}. The green point with its corresponding error-bar indicates the estimated value of $\Lambda_{1.4}$, provided by the GW170817 detection~\cite{Abbott-2019-X}. We notice that the EoSs with higher values of $\eta$ can not predict the observed value of $\Lambda_{1.4}$.

\begin{figure}[h]
\includegraphics[width=228pt,height=18pc]{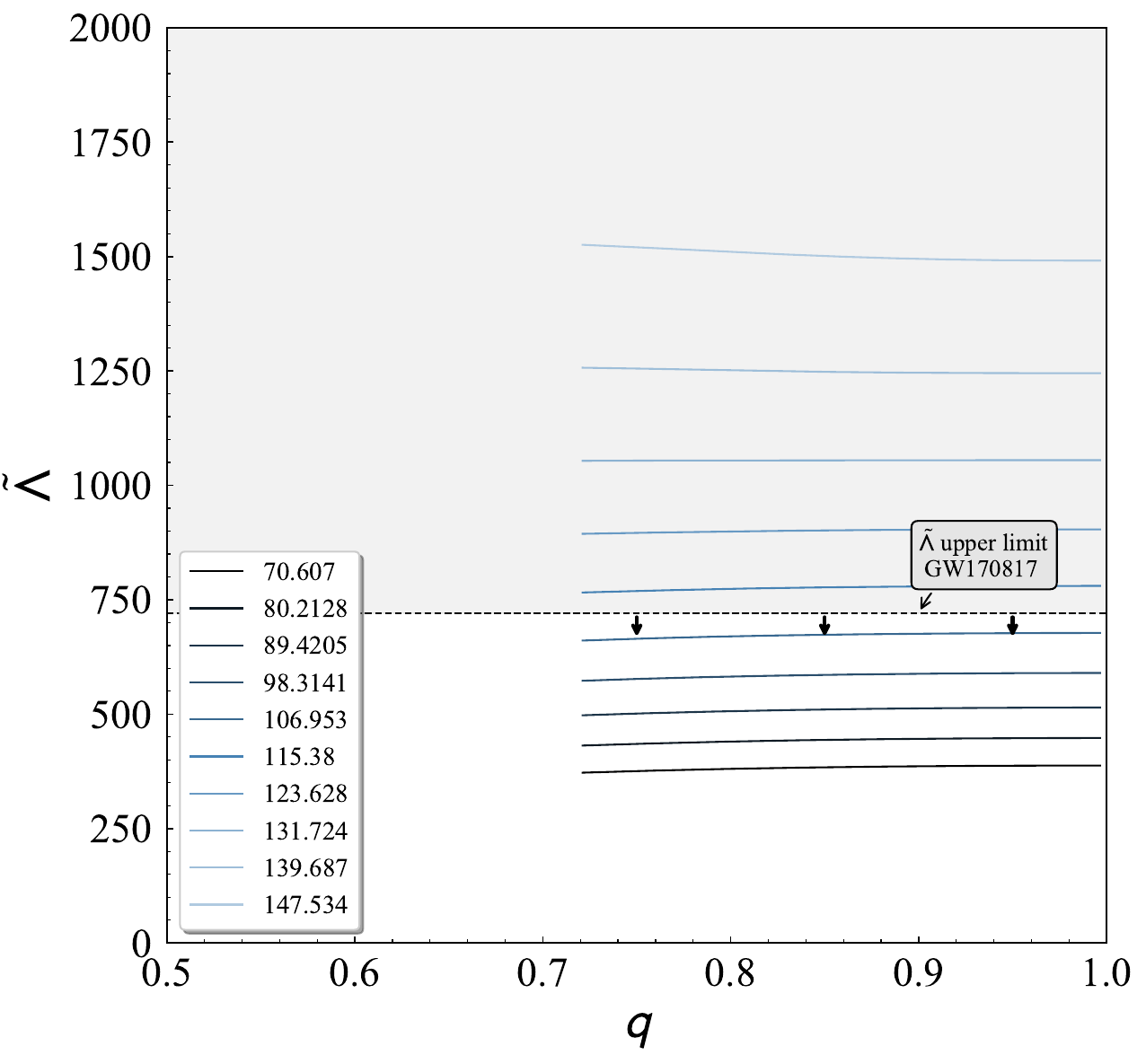}
\caption{ The effective tidal deformability $\tilde{\Lambda}$ vs the binary mass ratio $\mathrm{q}$ for all the cases of EoS, applied to the GW170817 event~\cite{Abbott-2019-X}. The gray region corresponds to the excluded values provided by LIGO.}
\label{fig:Ltilde-q}
\end{figure}

In Fig.~\ref{fig:Ltilde-q} we show the behavior of the EoSs, characterized by the $\mathrm{\eta}$ parameter, by applying them to the case of the GW170817 event. Specifically, we demonstrate the $\tilde{\Lambda}-q$ dependence, accompanied by the observational upper limit on $\tilde{\Lambda}$. As one can observe, the high values of $\eta$ lead to very high values of $\tilde{\Lambda}$, with a value of $\eta\approx110$ MeV to be the critical one for this specific event. In general, this behavior arises via the stiffness dependence of the EoS from the $\eta$ parameter.

\begin{figure}[h]
\includegraphics[width=228pt,height=18pc]{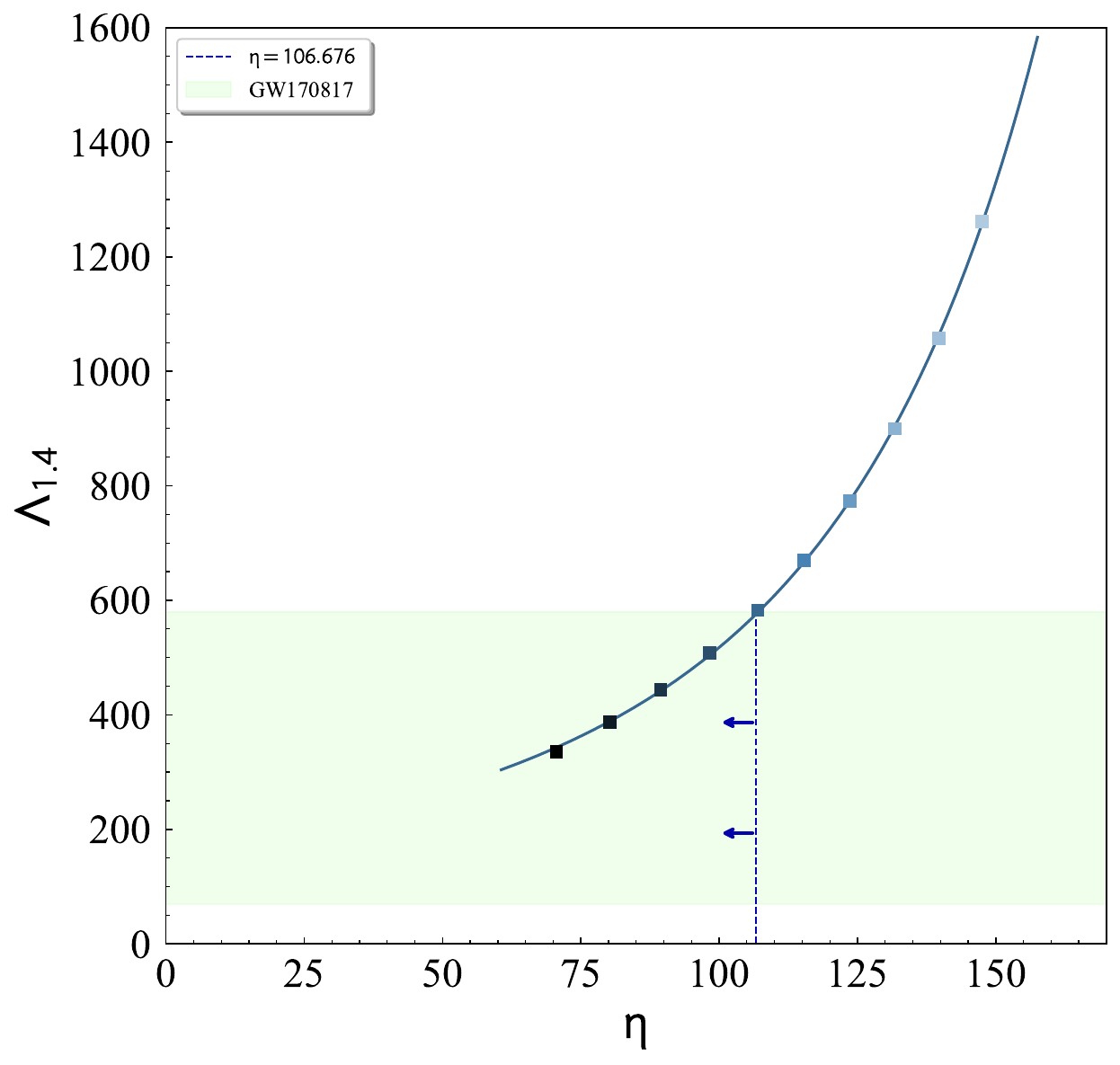}
\caption{ The tidal deformability $\Lambda_{1.4}$ of a $\mathrm{1.4\;M_\odot}$ neutron star related to the parameter $\mathrm{\eta}$. The green shaded area indicates the observational constraints from GW170817~\cite{Abbott-2019-X}.}
\label{fig:Lambda14-eta}
\end{figure}

For the need of examining further the dependence of the EoS from the $\eta$ parameter, we constructed Fig.~\ref{fig:Lambda14-eta}, in which the tidal deformability $\Lambda_{1.4}$ of a $\mathrm{1.4\;M_\odot}$ neutron star is studied as a relation of $\eta$. Each square point corresponds to the relevant EoS, characterized by the value of $\eta$. As  $\eta$ gets higher values, the color of points lightens. By applying the observational limits of $\Lambda_{1.4}$, provided by LIGO, we extracted an upper value of $\mathrm{\eta_{max}\simeq106.676}$ MeV so that all the EoSs with $\mathrm{\eta\leq\eta_{max}}$, indicated by the blue horizontal arrows in the figure, fulfill the observational constraints of GW170817. The blue color curve corresponds to a fitted formula, which in a good approximation is given in the following form
\begin{equation} 
\Lambda_{1.4}(\eta)=c_1\exp{(c_2^\eta)}, 
\end{equation}
where $c_1\simeq63.38614$ and $c_2\simeq1.00745$.

\begin{figure}[h]
\includegraphics[width=240pt,height=19pc]{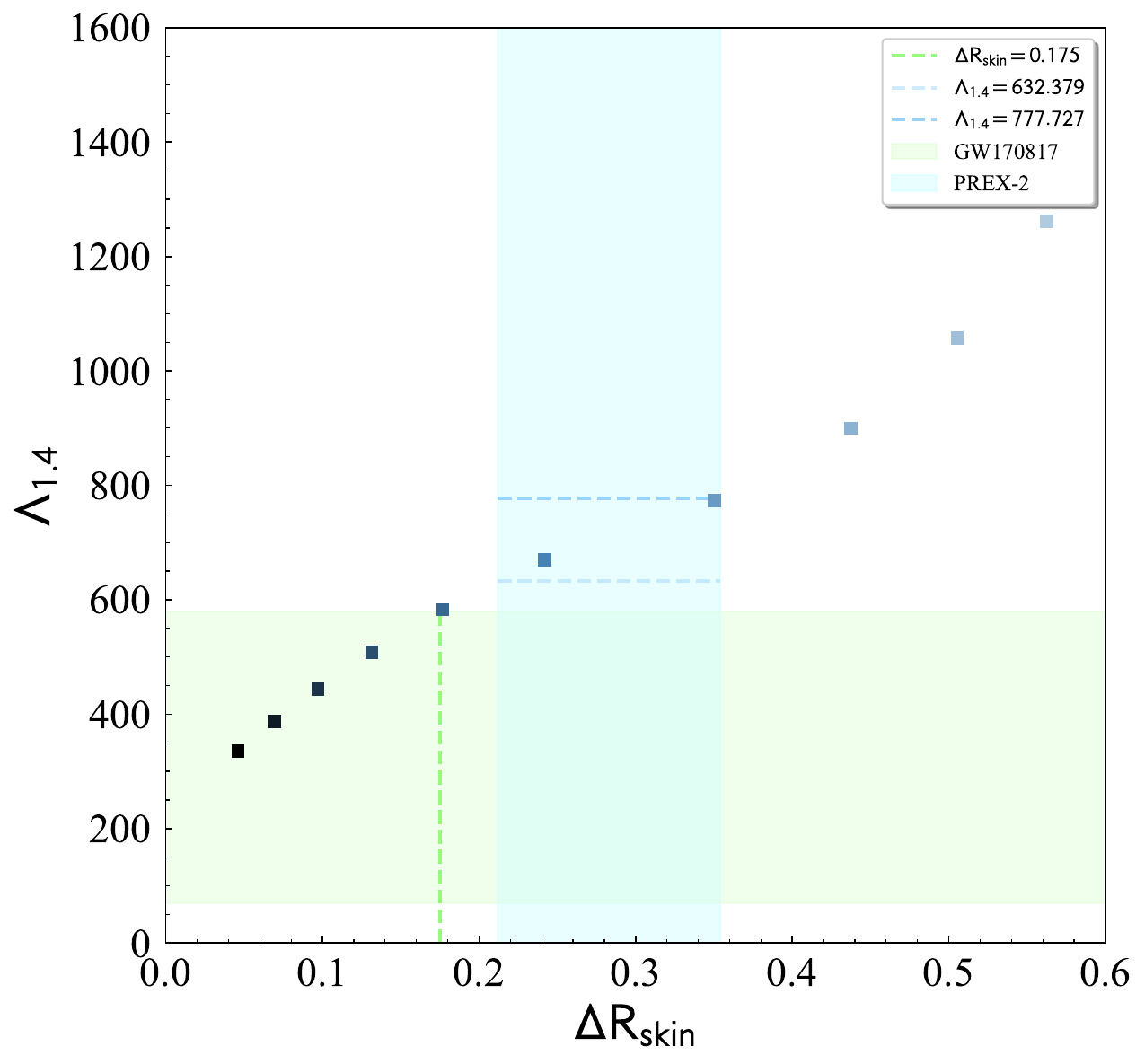}
\caption{ The tidal deformability $\Lambda_{1.4}$ of a $\mathrm{1.4\;M_\odot}$ neutron star related to the neutron skin $\mathrm{\Delta R_{skin}}$ (in fm). The green shaded area indicates the observational constraints on $\Lambda_{1.4}$ by GW170817~\cite{Abbott-2019-X}, while the blue one indicates the PREX-2 estimation for $\mathrm{\Delta R_{skin}}$~\cite{Adhikari-2021}.}
\label{fig:Lambda14-DRskin}
\end{figure}

Moving on to the neutron skin $\mathrm{\Delta R_{skin}}$, we studied its behavior related to $\Lambda_{1.4}$, aiming to extract further information from the observational constraints, as shown in Fig.~\ref{fig:Lambda14-DRskin}. The square points correspond to the relevant EoSs as described in the previous figure. The upper limit $\Lambda_{1.4}=580$ imposes an upper value for the neutron skin, $\mathrm{\Delta R_{skin}=0.175}$ (green dashed line), while the corresponding limits provided by PREX-2 are translated to the following acceptance region for $\Lambda_{1.4}\in[632.379,777.727]$ (blue horizontal dashed lines). The combination of these two constraints, originated from observational and experimental data, lead to different directions. The gravitational-wave origin leads to smaller values of the neutron skin, while the PREX-2 favors higher values. This contradiction arises from the softness of the EoS that the GW170817 imposes, while the PREX-2 requires a stiffer EoS.

\begin{figure}[ht]
\includegraphics[width=228pt,height=18pc]{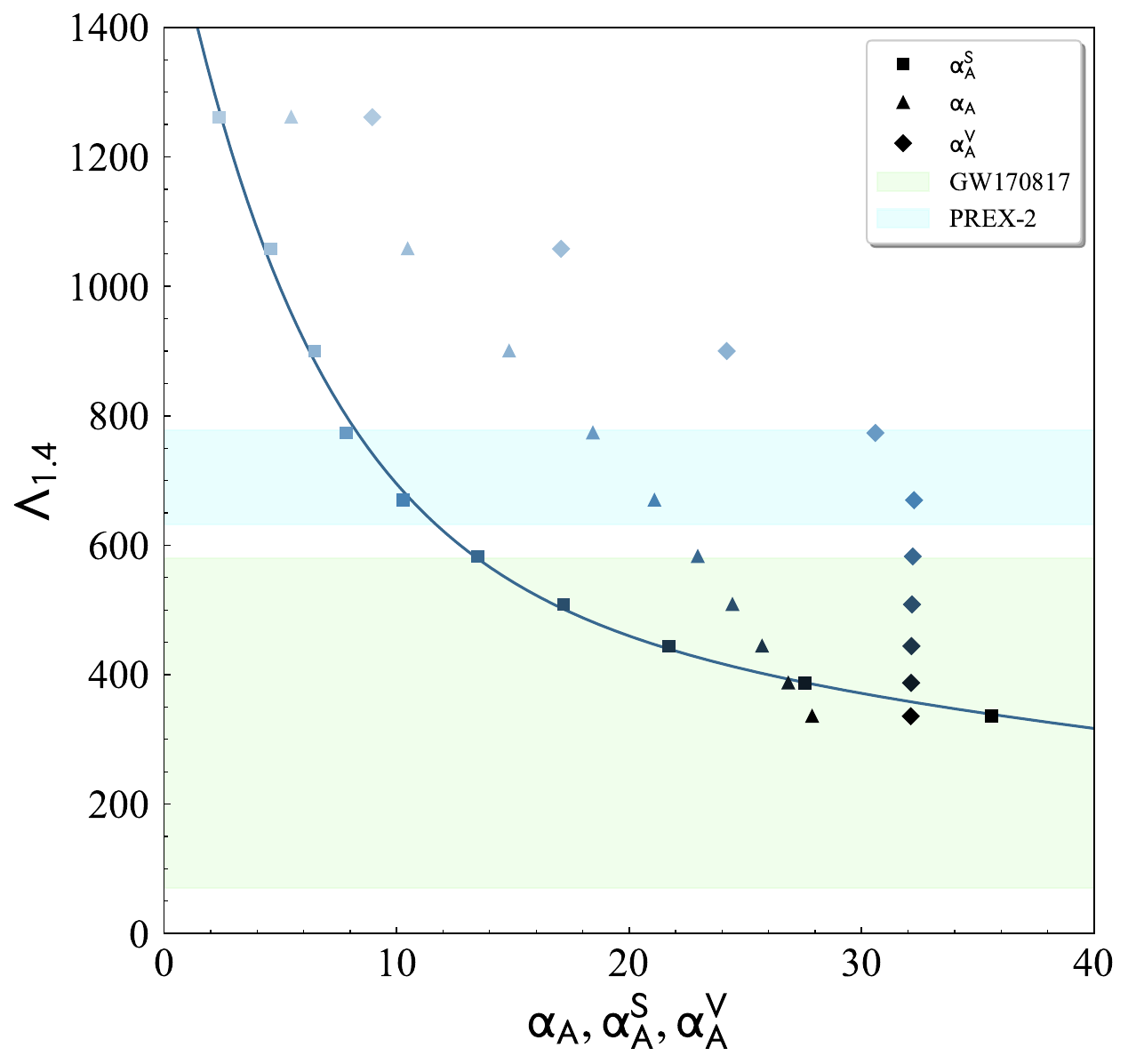}
\caption{The tidal deformability $\Lambda_{1.4}$ of a $\mathrm{1.4\;M_\odot}$ neutron star related to the asymmetry coefficient $\mathrm{\alpha_A}$ (in MeV) and the surface (volume) coefficient $\mathrm{\alpha_A^S}$ ($\mathrm{\alpha_A^V}$) (in MeV). The green shaded area indicates the observational constraints on $\Lambda_{1.4}$ by GW170817~\cite{Abbott-2019-X}, while the blue one indicates the corresponding PREX-2~\cite{Adhikari-2021} estimation for $\Lambda_{1.4}$.}
\label{fig:Lambda14-aparam}
\end{figure}

In order to take a deeper look into the microscopic parameters, we constructed  Fig.~\ref{fig:Lambda14-aparam}. In this kind of diagram we take advantage of the observational upper limit $\Lambda_{1.4}=580$ provided by GW170817 (green area), so that a lower limit on each parameter can be obtained. For the surface coefficient this limit corresponds to $\alpha_A^S\geq13.45837$. By applying the estimation region for $\Lambda_{1.4}$ that we extracted previously (provided by PREX-2 measurements on the neutron skin), the surface coefficient should lie inside $\alpha_A^S\in[8.23576,11.68261]$. The $\Lambda_{1.4}(\alpha_A^S)$ behavior can be described well by the following exponential formula

\begin{equation}
    \Lambda_{1.4}(\alpha_A^S)=c_3\exp(-\alpha_A^S/c_4)+c_5\exp(-\alpha_A^S/c_6)+c_7,
\end{equation}
where $c_3=1082.95$, $c_4=6.18075$, $c_5=552.26432$, $c_6=71.30309$, and $c_7=3.32712\times10^{-7}$. The distinct estimation values (originating from either observational data that we used)  reiterate for the other two microscopic parameters, $\alpha_A$ and $\alpha_A^V$, as one can observe from Fig.~\ref{fig:Lambda14-aparam}.

\section{Concluding Remarks}
The main conclusions of the present study can be summarized as follows

\begin{enumerate}

\item The neutron skin thickness and the coefficients  $a_A$, $a_A^S$ and  $a_A^V$ are sensitive on the parameter $\eta$ which characterizes the stiffness of the equation of state. The effect is very dramatic especially for high values of $\eta$ ($\eta>120 \ {\rm MeV}$) leading to abnormal values for these parameters.

\item  For the neutron skin thickness, in order to  be compatible with the results of the PREX-2 experiment, the range of the parameter $\eta$ must be in the range 
$110 \ {\rm MeV}  \lesssim \eta  \lesssim 125 \ {\rm MeV}$.

\item The effects of the parameter $\eta$ are also very pronounced in neutron stars properties. In particular,  the increasing of $\eta$ affects more the radius compared to the $\mathrm{M_{max}}$. The EoSs with the highest $\eta$ lie outside of the GW170817 observation (green shaded contours), while only the EoS with the lowest value of $\eta$ can predict the HESS observation.

\item    By applying the observational limits of $\Lambda_{1.4}$, provided by LIGO, we extracted an upper value of $\mathrm{\eta_{max}\simeq 106.676} \ {\rm MeV}$ so that all the EoSs with $\mathrm{\eta\leq\eta_{max}}$ fulfill the observational constraints of GW170817.

\item The combination of these two constraints, originated from observational and experimental data, lead to different directions. The gravitational-wave origin leads to smaller values of the neutron skin, while the PREX-2 favors higher values. This contradiction arises from the softness of the EoS that the GW170817 imposes, while the PREX-2 requires a stiffer EoS.

\item We present for a first time constraints  for the other three microscopic parameters, $\alpha_A$, $\alpha_A^S$, and $\alpha_A^V$ with the help of recent observations (related mainly with the tidal deformability). We conclude that if we define the tidal deformability or even more the radius of a neutron star more precisely, we will also be able to define even more precisely the range of these coefficients.

\end{enumerate}

A final comment is appropriate:  Although in the present study we use a simple model to simultaneously describe finite nuclei and neutron stars, the final results show that although the difference in their dimensions is huge  (from a few fm to a few Km), they can be directly connected due to the common  isovector dependence of their properties. Thus it would be reasonable to assume that future precise measurements of the properties of neutron stars will lead to a more precise determination of the microscopic structure of finite nuclei, especially those that are
neutron-rich, and vice versa. This work is an attempt in this direction.

\section*{Acknowledgments}
This work is supported by the Hellenic Foundation for Research and
Innovation (HFRI) under the 3rd Call for HFRI PhD Fellowships (Fellowship Number: 5657).


\end{document}